\documentclass[a4paper,11pt]{article}
\usepackage{jheppub}
\usepackage{chessfss}
\pdfoutput=1
\pdfpageattr {/Group << /S /Transparency /I true /CS /DeviceRGB>>}
\usepackage{hyperref}
\hypersetup{colorlinks=true,linkcolor=blue,citecolor=blue}
\usepackage{amsmath,amssymb,amsfonts}
\usepackage{bm}
\usepackage{float}
\usepackage{subfig}
\usepackage{scalefnt}
\usepackage{wrapfig}
\usepackage{fancybox}
\usepackage{tikz}
\usepackage{multirow}
\allowdisplaybreaks

\usetikzlibrary{arrows}
\tikzset{>=angle 60}
\tikzstyle{W}=[draw,circle,scale=.6]
\tikzstyle{B}=[draw,circle,fill=black,scale=.6]
\tikzstyle{H}=[draw,circle,fill=gray,scale=.6]
\tikzstyle{every picture}=[scale=.6,baseline=(current bounding box.south)]

\def\beq#1\eeq{\begin{align}#1\end{align}}

\newcommand{\be}{\begin{eqnarray}}
\newcommand{\ee}{\end{eqnarray}}
\newcommand{\bea}{\begin{eqnarray}}
\newcommand{\eea}{\end{eqnarray}}

\newcommand{\bn}{\begin{enumerate}}
\newcommand{\en}{\end{enumerate}}

\def\half{\frac{1}{2}}

\def\XXint#1#2#3{{\setbox0=\hbox{$#1{#2#3}{\int}$}
\vcenter{\hbox{$#2#3$}}\kern-.5\wd0}}

\newcommand{\RT}[4]{\ensuremath{\Theta\left[\begin{array}{c}#1\\#2\end{array}\right]\left(#3,#4\right)}}

\title{M5 branes and Theta Functions}

\begin{document}
\author{Babak Haghighat,}
\author{Rui Sun}
\affiliation{Yau Mathematical Sciences Center, Tsinghua University, Beijing, 100084, China}

\abstract{We propose quantum states for Little String Theories (LSTs) arising from M5 branes probing A- and D-type singularities. This extends Witten's picture of M5 brane partition functions as theta functions to this more general setup. Compactifying the world-volume of the five-branes on a two-torus, we find that the corresponding theta functions are sections of line bundles over complex 4-tori. This formalism allows us to derive Seiberg-Witten curves for the resulting four-dimensional theories. Along the way, we prove a duality for LSTs observed by Iqbal, Hohenegger and Rey.}

\maketitle

\section{Introduction}

Since their discovery, M5 branes have played an important role in modern mathematical physics and have led to surprising new insights into string theory and superconformal quantum field theories in various dimensions. Put on various backgrounds, they give rise to intriguing nD-(6-n)D correspondences, the most well-known of which are the 2d-4d correspondence of \cite{Alday:2009aq} and the 3d-3d correspondence of \cite{Dimofte:2011ju} among others. In these correspondences the six-dimensional world-volume of the five-brane is compactified on a Riemann surface, a three-manifold or a four-manifold and the physics of the resulting gauge theories in lower dimensions is reflected in many properties of the geometry and topology of these manifolds. One surprising aspect of these results is that they are valid despite the fact that there is no action available for the M5 brane theory, although some indirect derivations have been found (see for example \cite{Cordova:2013cea,Cordova:2016cmu}).

In order to circumvent the fact that there is no Lagrangian description available, Witten proposed in \cite{Witten:1996hc} to view partition functions of M5 branes as vectors in a certain quantum Hilbert space. This Hilbert space arises by realizing the M5 brane world-volume as a boundary of a seven-dimensional theory whose path integral with a suitable boundary condition gives the corresponding state. This is similar to the path integral of Chern-Simons theory on a three-manifold with boundary a Riemann surface $\Sigma$ where the quantum Hilbert space is the space of sections of a certain line bundle over the Jacobian of $\Sigma$. In the case of the M5 brane theory these sections are theta functions over the intermediate Jacobian of the world-volume manifold as will be reviewed in more detail in Section \ref{sec:M5}. Witten's construction was later generalized in \cite{Belov:2006jd} to the case beyond spin manifolds. Extending these results, reference \cite{Monnier:2017klz} argues how conformal block of 6d $(2,0)$ theories of ADE type can be obtained from the seven-dimensional viewpoint. 

One might ask whether Witten's approach can be generalized to the case of 6d $(1,0)$ theories where there is equally no action principle available. Such theories have recently enjoyed a resurgence due to the discovery of a vast geometric classification through F-theory \cite{Heckman:2013pva}. First steps towards this direction were undertaken in \cite{DelZotto:2015isa} where the authors constructed the defect groups for various 6d SCFTs. 

In the current paper we want to further generalize to the case of 6d $(1,0)$ Little String Theories (LSTs). Here it turns out that symmetry enhancement leads to surprising new structures involving Riemann theta functions on complex 4-tori and their various properties. Again the M5 brane case is a guiding principle as we look at 6d theories arising from M5 branes probing ADE singularities. In particular, the backgrounds we look at are provided by M5 branes on a certain limit of the Omega background \cite{Nekrasov:2010ka} probing $S^1 \times \mathbb{C}^2/\Gamma_{A,D}$ where the singularity is either of A-type or of D-type. Such theories are then labeled by two integers, one of them being the number of M5 branes and the other the degree of the singularity. One novel viewpoint presented in this paper is that quantum states of the M5 brane theory in this background, given by theta functions, can be interpreted as quantum vacua of the resulting 4d $\mathcal{N}=2$ theory after torus compactification. This way we obtain a new derivation for Seiberg-Witten curves of such theories using theta functions. Moreover, we find that the process of gluing two M5 brane theories together to obtain a third theory with higher rank and higher singularity degree defines an operator product expansion at the level of quantum states which is nothing else than Riemann's addition formula for theta functions.
Armed with these new insights we then proceed to prove the duality web for A-type LSTs observed in \cite{Hohenegger:2016yuv} at the level of the Seiberg-Witten curve of such theories. In our context this duality web turns out to be simply obtained by unimodular transformations which keep the underlying lattice of our theta functions invariant. We then proceed to the D-type case where we obtain the corresponding theta functions in terms of a $\mathbb{Z}_2$ orbifold construction. Here the operator product described above becomes important as M5 branes probing D-type singularities fractionate into two $\half\textrm{M5}$ branes whose wave-functions will pick up a $\pm$-sign under the $\mathbb{Z}_2$ action. Thus invariant states only appear at degree 2.

The organization of the paper is as follows. In Section \ref{sec:M5} we review Witten's construction of M5 brane partition functions as quantum states and specialize to the case of the Omega background. We then proceed to generalize this construction to the case of LSTs in Section \ref{sec:LST} and argue how the corresponding quantum states are theta functions over a 4-torus. In Section \ref{sec:theta} we give a detailed overview of theta functions as sections of line bundles over abelian varieties and introduce many of their properties in a rather detailed manner. This discussion includes Sections \ref{sec:basics} and \ref{sec:addition} which give a count of the number of independent sections, characteristics, transformation properties under lattice shifts and under modular transformations, as well as the addition formula. Next, in Section \ref{sec:M5A} we turn to the case of M5 branes probing an A-type singularity and derive the corresponding Seiberg-Witten curve for this case and show its symmetries. Finally, in Section \ref{sec:M5D}, we turn to the D-type case and first give a heuristic description of the setup before turning to the mathematical details involving even and odd theta functions and their products. We close with a discussion Section giving an overview of future lines of research.

\section{Quantum States of LSTs}

In this Section we describe how LSTs give rise to a quantum Hilbert space of vacua. To this end we start by reviewing Witten's construction of the effective action of M5 branes in M-theory where he introduces the notion of M5 brane states as sections of a certain line bundle \cite{Witten:1996hc}. We then proceed to generalize this construction to the case of LSTs where our primary example is the case of M5 branes probing A-type singularities. This construction then allows us to describe the moduli space of vacua of LSTs compactified on a two-torus down to a 4d $\mathcal{N}=2$ theory elegantly in terms of Riemann-Theta-Functions. 

\subsection{Review of Witten's construction}
\label{sec:M5}

In \cite{Witten:1996hc} Witten showed that partition functions of M5 branes on a six-manifold $W$ can be  understood as wave-functions which depend on the value of a certain background gauge three-form $C$. The fact that the two-form on the M5-brane worldvolume is chiral (or self-dual) implies that these wave-functions are holomorphic in a certain sense. This holomorphy plus gauge-invariance under complexified gauge transformations then together imply that $C$ defines a point in $H^3(W,\mathbb{R})$. Dividing further by ``big gauge transformations" we then see that this space descends to $J_W = H^3(W,\mathbb{R})/H^3(W,\mathbb{Z})$, which is known as the \textit{intermediate Jacobian} of $W$. Thus partition functions or states of the M5 brane are sections of a certain line bundle $L$ over the torus $J_W$. These \textit{states} can then be thought of as arising from a path-integral of a seven-dimensional theory with boundary $W$ similar to the path-integral of Chern-Simons theory on a three-manifold with boundary a Riemann surface $\Sigma$. In fact, Witten argues that for a single M5 brane a unique section of $L$ is singled out. 

In the following we want to focus on the case where $W = \mathbb{R}^4 \times \mathbb{T}_{\tau}$ where $\mathbb{T}_{\tau}$ is a two-torus with complex structure $\tau$, i.e.
\begin{equation}
	\mathbb{T}_{\tau} \equiv \mathbb{C}/\left(\mathbb{Z} \oplus \tau \mathbb{Z}\right).
\end{equation}
Suppose that for simplicity $\mathbb{T}_{\tau}$ is just the direct product of two circles, i.e. $\mathbb{T}_{\tau} = S^1 \times \widetilde{S}^1$. In this case one can write $H^3(W,\mathbb{Z}) = A \oplus B$, where\footnote{As the cohomology of $\mathbb{R}^4$ is trivial, we use here a regularized version where we blow-up $\mathbb{R}^4 = \mathbb{C}^2$ at the origin and denote the resulting space by $\widehat{\mathbb{C}^2}$. Then we define $H^2(\mathbb{R}^4,\mathbb{Z}) \equiv \lim_{\textrm{Vol}(E) \rightarrow 0}H^2(\widehat{\mathbb{C}^2},\mathbb{Z})=\mathbb{Z}$, where $E$ is the exceptional $\mathbb{P}^1$. Such a regularization scheme is not uncommon and has already been used in similar contexts before, see e.g. \cite{Nakajima:2003pg}.}
\begin{equation}
	A = H^2(\mathbb{R}^4,\mathbb{Z}) \otimes H^1(S^1,\mathbb{Z}), \quad B = H^2(\mathbb{R}^4,\mathbb{Z}) \otimes H^1(\widetilde{S}^1,\mathbb{Z}).
\end{equation}
This shows that the Jacobian is just the two-torus we started with, i.e.
\begin{equation}
	J_W = H^3(\mathbb{R}^4 \times \mathbb{T}_{\tau},\mathbb{R})/H^3(\mathbb{R}^4\times \mathbb{T}_{\tau},\mathbb{Z}) = \mathbb{T}_{\tau}.
\end{equation}
Therefore, holomorphic sections of line bundles over $J_W$ are just ordinary theta-functions. As argued in \cite{Witten:1998wy,Witten:2009at}, this story can be generalized to the case of multiple M5 branes as follows. First of all, notice that the case of multiple M5 branes is just a special case of a more general characterization of $(2,0)$ theories by a simply-laced Lie Group $G$ (arising from Type IIB string theory on an ADE singularity) and corresponds to the A-type case. In general, if $G$ is simple, simply-laced, and simply-connected, its center is $\mathcal{Z} = \Gamma^{\vee}/{\Gamma}$ where $\Gamma$ and $\Gamma^{\vee}$ denote the root lattice and its dual. Then the quadratic form on $\Gamma$ leads to a perfect pairing 
\begin{equation}
	\mathcal{Z} \times \mathcal{Z} \rightarrow \mathbb{R}/\mathbb{Z} = U(1).
\end{equation}
Together with Poincar\'e duality, we then get a perfect pairing $H^3(W,\mathcal{Z}) \times H^3(W,\mathcal{Z}) \rightarrow U(1)$. Taking $W=\mathbb{R}^4 \times \mathbb{T}_{\tau}$, we again get a factorization $H^3(W,\mathcal{Z}) = A \oplus B$, where $A$ and $B$ are given as above with the replacement $\mathbb{Z} \mapsto \mathcal{Z}$. It can be shown then that the space of wave-functions has a basis $\psi_a$, $a \in A$. Exchanging the roles of $A$ and $B$, one finds a second basis $\widetilde{\psi}_b, b \in B$. These two bases are related by
\begin{equation} \label{eq:S-duality}
	\widetilde{\psi}_b = \psi_b(-1/\tau,z/\tau) = C \sum_{a\in H^2(\mathbb{R}^4,\mathcal{Z})} \exp(2\pi i (b,a)) \psi_a(\tau,z),
\end{equation}
where $C$ is a ($z$-dependent) constant, $z$ a point in  $\mathbb{T}_{\tau}$, and we write $\exp(2\pi i (a,b))$ for the perfect pairing $H^2(\mathbb{R}^4,\mathcal{Z}) \times H^2(\mathbb{R}^4,\mathcal{Z}) \rightarrow U(1)$. Now observe that $H^2(\mathbb{R}^4,\mathcal{Z}) \cong \mathcal{Z}$ and thus $a$ runs over all elements in $\mathcal{Z}$. It is worth taking a closer look at $\mathcal{Z}$ at this point. $\mathcal{Z}$ is an abelian group which can be identified with the center of $G$. In fact, it is a discrete group and in the case of ADE groups with Lie Algebra denoted by $\frak{g}$ it can be expressed through the cyclic groups (see the Appendix of \cite{Nekrasov:2012xe} for more details) shown in Table \ref{tab:cyclic}.
\begin{table}
\centering
\begin{tabular}{c|c||c}
\hline
$G$ & $z_{\frak{g}}$ & $l_i, i=1,\ldots, z_{\frak{g}}$ \\
\hline 
$A_r$ & $1$ & $r+1$ \\
$D_{2s}$ & $2$ & $2,2$ \\
$D_{2s+1}$ & $1$ & $4$\\
$E_6$ & $1$ & $3$ \\
$E_7$ & $1$ & $2$ \\
$E_8$ & $0$ & ~ \\
\hline
\end{tabular}
\caption{The number of cyclic factors in $\mathcal{Z}$ with their orders.}
\label{tab:cyclic}
\end{table}
The isomorphism between $\mathcal{Z}$ and a product of cyclic groups is then given as follows
\begin{equation}
	\mathcal{Z} \cong \bigotimes_{i=1}^{z_{\frak{g}}} \mathbb{Z}/l_i \mathbb{Z}
\end{equation}
for some $z_{\frak{g}}$, which is equal to $0$, $1$, or $2$ as shown in the table. Let us apply this reasoning to $A_{M-1}$-type $(2,0)$ theories. Then we see that the sum in (\ref{eq:S-duality}) becomes a sum over elements of $H^2(\mathbb{R}^4,\mathcal{Z}) = \mathbb{Z}_{M}$.  Thus we recognize in (\ref{eq:S-duality}) the transformation property of degree $M$ theta-functions under $S$-transformations, where we refer to \cite{Lange} and Section \ref{sec:theta} for more details. Such theta functions are sections of a line bundle $L^{M}$ which is the $M$th tensor power of a primitive line bundle $L$ such that we have
\begin{equation}
	\textrm{dim}_{\mathbb{C}}H^0(L) = 1, \quad \textrm{dim}_{\mathbb{C}}H^0(L^{M}) = M.
\end{equation}
Such sections are given in terms of linearly independent theta-functions $\Theta[\alpha]$ where $\alpha$ is running from $0$ to $M-1$. Let us next generalize this construction to $\mathcal{N}=(1,0)$ LSTs where our primary example will be the little string theory arising from $M$ M5 branes probing an $A_{N-1}$-singularity of the form $\mathbb{C}^2/\mathbb{Z}_N$.

\subsection{LST states as sections of Line Bundles}
\label{sec:LST}

We now turn to the case of LSTs and again put the theories on $\mathbb{R}^4 \times \mathbb{T}_{\tau}$. It turns out that in this case the symmetry group is enhanced as compared to the $S$-duality symmetry discussed in the previous paragraph. The reason is that Little String Theories have an intrinsic (string)-scale and enjoy T-duality invariance.  Let's see how this comes about in more detail. Consider $M$ M5 branes probing an ADE singularity in M-theory along a circle $S^1_{\perp}$ with radius $R \sim \rho$. Let us denote this theory by $\mathcal{T}_{M,\frak{g}}$ where $\frak{g}$ is the Lie Algebra corresponding to the ADE singularity. Sending $R \rightarrow \infty$ gives a 6d SCFT which is one of the conformal matter theories introduced in \cite{DelZotto:2014hpa}. But we don't want to do this here and rather want to keep $\rho$ finite. As discussed in \cite{Ohmori:2015pia}, in order to arrive at the T-dual theory, one can first take the M-theory circle to be one of the circles in $\mathbb{T}_{\tau}$ and compactify along it. This gives a Type IIA description where $r$ D4 branes are probing the singularity. One then T-dualizes along the circle $S^1_{\perp}$ giving rise to D5 branes probing the ADE singularity. However, something has changed now: the world-volume theory of the resulting T-dual 6d LST is not $\mathbb{R}^4 \times \mathbb{T}_{\tau}$ but rather $\mathbb{R}^4 \times \mathbb{T}_{\rho}$, where $\mathbb{T}_{\rho}$ is a two-torus with complex structure $\rho$. To see the implications of this, let's focus on the case of an $A$-type singularity $\mathbb{C}^2/\mathbb{Z}_N$ and uplift the T-dual theory to M-theory. This then gives $N$ M5 branes probing a $\mathbb{C}^2/\mathbb{Z}_M$ singularity, namely the theory $\mathcal{T}_{N,M}$\footnote{In the following, whenever the Lie Algebra $\frak{g}$ is of $A_{M-1}$-type, we write $\mathcal{T}_{N,M}$ instead of $\mathcal{T}_{N,\frak{g}}$.}. So the roles of $M$ and $N$ have been switched. This is known as fiber-base duality in the corresponding F-theory construction \cite{Bhardwaj:2015oru}. The theory $\mathcal{T}_{M,N}$ arises in F-theory by compactification on a doubly-elliptic non-compact Calabi-Yau 3-fold such that the elliptic fiber degenerates according to an affine $A_{N-1}$-singularity and the base elliptic curve is affine $A_{M-1}$. The T-dual theory is the theory with the roles of fiber and base switched. The picture presented here does not assume that the M5 branes form a stack and we do allow for moduli corresponding to a finite separation. This will become important later on when we come to the description of the Seiberg-Witten curve which will depend on all moduli. 

It is now clear that in the full theory the symmetries are generated by the symmetries of each of the two two-tori $\mathbb{T}_{\tau}$ and $\mathbb{T}_{\rho}$, denoted by S-dualities\footnote{These generate the S-dualities of resulting 4d theories.}, as well as the T-duality symmetry which exchanges the two two-tori:
\begin{equation}
	\left(\begin{array}{cc}\sigma_1 & 0 \\ 0 & \sigma_1 \end{array}\right), \quad \textrm{ where } \quad  \sigma_1 = \left(\begin{array}{cc}0 & 1\\ 1 & 0\end{array}\right).
\end{equation}
Together these symmetries generate the group $Sp(4,\mathbb{Z})$ which is the symmetry group of a $4$-torus $\mathbb{T}_{\tau} \times \mathbb{T}_{\rho}$. In fact, in the case of the $\mathcal{T}_{M,N}$ theories the torus is an abelian surface
\begin{equation}
	\mathbb{T}_{\Omega} \equiv \mathbb{C}^2/\left(\mathbb{Z}^2 \oplus \Omega \mathbb{Z}^2\right), \quad \Omega = \left(\begin{array}{cc}\tau & \sigma \\ \sigma & \rho \end{array}\right),
\end{equation}
where $\sigma$ is the mass-deformation of the $A$-type theory \cite{Haghighat:2017vch}. For the following discussion we will set $\sigma = 0$ for simplicity but it should be clear that all statements will hold for non-zero $\sigma$ as well as one can carefully check. Note that the self-dual 3-form of M5 branes can now couple to background 3-forms $C$ which take values in the Jacobian of this abelian surface. This is easy to see, as in the theory $\mathcal{T}_{M,N}$ the values of $C$ are characterized by points on $\mathbb{T}_{\tau}$ and in the T-dual theory $\mathcal{T}_{N,M}$ they are characterized by points on $\mathbb{T}_{\rho}$. Thus altogether we get for the intermediate Jacobian of the Little String Theory:
\begin{equation}
	J^{\textrm{LST}}_{W} \cong H^1(\mathbb{T}_{\tau}\times \mathbb{T}_{\rho},\mathbb{R})/H^1(\mathbb{T}_{\tau}\times \mathbb{T}_{\rho},\mathbb{Z}) = \mathbb{T}_{\Omega}.
\end{equation}
This also shows that LST states will be described by sections of line bundles over $\mathbb{T}_{\Omega}$. Let us describe in a bit more detail what properties such sections should have. First of all, note that restriction of such a line bundle to one of the tori, let's say $\mathbb{T}_{\tau}$, should give back the story discussed in the previous subsection. Namely, for $M$ M5 branes the restriction of the line bundle should give $M$ sections. Similarly, the restriction to the other torus should give $N$ sections. Thus sections will be labeled by two numbers $\alpha$ and $\beta$ and we have
\begin{equation}
	\textrm{dim}_{\mathbb{C}}(L) = M N, 
\end{equation}
with basis given by theta functions
\begin{equation}
	\RT{(\alpha,\beta)}{}{\Omega}{\vec{z}}, \quad \alpha = 0, \ldots, M-1,~~\beta=0, \ldots, N-1,
\end{equation}
where $\vec{z} \in \mathbb{C}^2$. We will give precise definitions of these theta-functions in Section \ref{sec:theta} where we will also discuss their properties. Let us next see what implications this gives for the quantum moduli space of vacua.

\subsection{Quantum Moduli Spaces of Vacua}
\label{sec:Qstates}

In the previous subsection we have seen that for the theory $\mathcal{T}_{M,N}$ each state corresponds to a theta function. In fact, the theta functions form a $\mathbb{C}$-basis for the Hilbert space and any basis vector is given up to a multiplicative pre-factor
\begin{equation}
	a_{\alpha,\beta}\RT{(\alpha,\beta)}{}{\Omega}{\vec{z}}, \quad a_{\alpha,\beta} \in \mathbb{C}.
\end{equation}
An arbitrary vacuum state is then given in terms of a linear combination of the form 
\begin{equation} \label{eq:section}
	\sum_{\alpha,\beta} a_{\alpha,\beta} \RT{(\alpha,\beta)}{}{\Omega}{\vec{z}}.
\end{equation}
This will then be an arbitrary section of the line bundle $L_{M,N}$. Therefore, the moduli space of vacua of the theory is the parameter space given by the $a_{\alpha,\beta}$ up to the action of the symplectic group $Sp(4,\mathbb{Z})$ which will be discussed in Section \ref{sec:theta}. Furthermore, note that a section given by (\ref{eq:section}) uniquely corresponds to a divisor in $\mathbb{T}_{\Omega}$ specified by
\begin{equation} \label{eq:SWtheta}
	\sum_{\alpha,\beta} a_{\alpha,\beta} \RT{(\alpha,\beta)}{}{\Omega}{\vec{z}} = 0.
\end{equation}
This is easy to see as the left-hand side is a section and transforms up to a multiplicative factor under ``large gauge transformation"  given by the action of $H^1(\mathbb{T}_{\Omega},\mathbb{Z})$. Thus the only way we can form an equation is to set the right-hand side to zero to make it invariant. Now such a divisor is naturally of complex co-dimension one and thus a Riemann surface $\Sigma \subset \mathbb{T}_{\Omega}$. We now see the emergence of the Seiberg-Witten curve of the four-dimensional $\mathcal{N}=2$ theory obtained by torus compactification of our 6d LST.  We would like to note here that the representation (\ref{eq:SWtheta}) for the Seiberg-Witten curve was derived in the past, using different methods, by Braden and Hollowood \cite{Braden:2003gv}\footnote{In \cite{Braden:2003gv}, the authors focus mainly on the case $M=1$.}. The wave-function viewpoint adopted in our paper will allow us to go beyond this result and also obtain expressions for LSTs arising from M5 branes probing $D$-type singularities. The goal of the next Section will be to sharpen our reasoning by utilizing the mathematical theory of Riemann-Theta functions and apply this machinery to deduce new results.

\section{Theta Functions}
\label{sec:theta}

In this Section we want to study line bundles on the Jacobian $J$ of the Abelian Surface $\mathbb{T}_{\Omega}$. We show how properties of sections of such line bundles interpreted as M5 brane states allow us to deduce the moduli spaces of vacua for M5 branes probing A-type and D-type singularities together with their duality frames.

\subsection{Some basics}
\label{sec:basics}

To begin the discussion, we start from a generalized setting where $J = \mathbb{R}^{2n}/\Lambda$, where $\Lambda$ is a rank $2n$ lattice in $\mathbb{R}^{2n}$. Then we call an element $\omega \in H^2(J,\mathbb{Z})$ a \textit{principal polarization} when
\begin{equation}
	\int_J \frac{\omega^n}{n!} = 1.
\end{equation}
An example of such an $\omega$ is as follows. Let $x^i$, $y_j$, $i,j=1, \ldots, n$ be coordinates on $\mathbb{R}^{2n}$ such that $\Lambda$ is spanned by unit vectors $e_i$  and $f^j$ in the $x^i$ and $y_j$ coordinates, respectively. Then $\omega = \sum_i dx^i \wedge dy_i$ defines a principal polarization and we also write
\begin{equation}
	\omega = \omega(e_i, f^j) = -\omega(f^j,e_i) =  \left(\begin{array}{cc} 0 & \mathbf{1}_n \\ - \mathbf{1}_n & 0 \end{array}\right).
\end{equation}
Conversely, any translation-invariant two-form $\omega$ representing a principal polarization can be put in such a form by a suitable choice of coordinates. In general, we can have more complicated polarizations where $\omega = \sum_i d_i dx^i \wedge dy_i$ for some integers $d_i$. Then there exists a unimodular transformation of the lattice $\Lambda$ such that $d_i \geq 0$ satisfying $d_i | d_{i+1}$ \cite{Lange}. In such a case, $\omega$ written as a matrix takes the form
\begin{equation} \label{eq:omega}
	\omega = \left(\begin{array}{cc} 0 & D \\ - D & 0 \end{array}\right),
\end{equation}
where $D = \textrm{diag}(d_1,\ldots,d_n)$. A line bundle $L$ on $J$ with the projection
\begin{equation}
	\pi :  L \rightarrow J,
\end{equation}
is topologically up to isomorphism uniquely specified in terms of its first Chern class $c_1(L) = \omega$. To give a more complete definition of the line bundle $L$ we also need to fix its $U(1)$ connection which is a $U(1)$ gauge field on $J$ denoted by $A$ with the property that $F=dA$ equals $2\pi \omega$. To fix $A$ we must give, in addition to the curvature, the holonomies around noncontractible cycles in $J$. To do that, we proceed as follows. Specifying $\omega$ as in (\ref{eq:omega}) leads to a decomposition for $L$ given by 
\begin{equation}
	\Lambda = \Lambda_1 \oplus \Lambda_2
\end{equation}
with $\Lambda_1 = \langle \lambda_1,\ldots, \lambda_n\rangle$ and $\Lambda_2 = \langle \mu_1,\ldots,\mu_n\rangle$. This induces a decomposition for $\mathbb{R}^{2n}$:
\begin{equation} \label{eq:decomp}
	\mathbb{R}^{2n} = V_1 \oplus V_2,
\end{equation}
such that $\Lambda_{\nu} = V_{\nu} \cap \Lambda$ for $\nu = 1,2$. For each $\lambda \in \Lambda$ we can now specify a closed curve $C(\lambda)$ in $J$ which is given by a straight line from the origin of $\mathbb{R}^{2n}$ to $\lambda$. Let $\chi(\lambda) = \exp\left(i \int_{C(\lambda)} A\right)$ be the holonomy of $A$ around $C(\lambda)$. $A$ will be completely fixed once the $\chi(\lambda)$ are given. The $\chi$'s are called \textit{semicharacters} on $\Lambda$ and are constrained as follows. If $\lambda$ and $\mu$ are any two lattice points, then (see \cite{Witten:1996hc})
\begin{equation}
	\chi(\lambda + \mu) = \chi(\lambda) \chi(\mu) (-1)^{\omega(\lambda,\mu)}.
\end{equation}
Thus a line bundle $L$ on $J$ is uniquely specified by a pair $\omega$ and $\chi$. In fact, there is a one-to-one correspondence between any symplectic form $\omega$ and hermitian forms $H$ where one defines $\omega = \textrm{Im}{H}$. Conversely, given a form $\omega$, we can construct a hermitian form out of it by defining
\begin{equation}
	H(v,w) = \omega(i v, w) + i \omega(v,w).
\end{equation}
Given this correspondence, we henceforth will parametrize a line bundle $L$ by $L = (H,\chi)$. The tensor product of two line bundles $L_1$ and $L_2$ can be then expressed as
\begin{equation}
	L_1 \otimes L_2 = (H_1 + H_2, \chi_1 \chi_2).
\end{equation}

\subsubsection*{Sections of $L$}
In the following we want to give an explicit representation of sections of a given line bundle $L=(H,\chi)$. First of all, we need to figure out the dimension of the space of sections $H^0(J,L)$. To this end, note that the connection $A$ determines a complex structure on $L$. The index of the $\overline{\partial}$ operator on $J$, with values in $L$, is
\begin{equation}
	\sum_{i=0}^{\textrm{dim}_{\mathbb{C}}J} (-1)^ i \textrm{dim} H^i(J,L) = \int_J e^{c_1(L)} \textrm{Td}(J) = \int_J e^{\omega} = d_1 \cdot d_2 \cdots d_n.
\end{equation}
Since $\omega$ is positive\footnote{We restrict ourselves here to an ample line bundle.}, the cohomology $H^i(J,L) = 0$ for $i > 0$, so the index formula gives us
\begin{equation}
	\textrm{dim}_{\mathbb{C}} H^0(J,L) = d_1 \cdot d_2 \cdots d_n.
\end{equation}
Note that in the case of principal polarization there is only one section up to multiplication while for non-principal polarizations there can be many. Let us construct these sections which we shall denote by $\Theta^L_i$ with $i=1, \ldots, d_1 \cdots d_n$. To this end, we note that the decomposition (\ref{eq:decomp}) leads to an explicit description of all line bundles L. Let us see how this comes about. Define a map $\chi_0 : \mathbb{R}^{2n} \rightarrow \mathbb{C}^*$ by
\begin{equation}
	\chi_0(z) = \exp(\pi i \omega(z_1,z_2)),
\end{equation}
where $z = z_1 + z_2$ with $z_{\nu} \in V_{\nu}$. It can be easily seen that $\left.\chi_0\right|_{\Lambda}$ is a semicharacter for $H$. Define a corresponding line bundle $L_0$ given by $L_0 = (H,\chi_0)$. Then it can be shown that all line bundles $L = (H,\chi)$ can be constructed out of $L_0$ \cite{Lange}. Namely, for every $L$ with the same $H$ there is a point $c \in \mathbb{R}^{2n}$, uniquely determined up to translation by elements of $\Lambda(L)$, such that $L \cong t_{\bar{c}}^* L_0$ where $t_{\bar{z}}$ denotes the translation operator by $z$. Equivalently, this gives $\chi = \chi_0 \exp(2\pi i \omega(c,\cdot))$. In the above, $\Lambda(L)$ denotes the lattice
\begin{equation}
	\Lambda(L) = \{ z \in \mathbb{R}^{2n} | \omega(z,\Lambda) \subset \mathbb{Z} \}.
\end{equation}
$c$ is called the \textit{characteristic} of $L$. Sections of $L$, or in other words elements of $H^0(L)$, can be identified with the set of holomorphic functions $\Theta : \mathbb{R}^{2n} \rightarrow \mathbb{C}$ satisfying
\begin{equation} \label{eq:shiftsym}
	\Theta(z + \lambda) = e_L(\lambda,z) \Theta(z),
\end{equation}
where $e_L(\lambda,z)$ is the \textit{classical factor of automorphy} and is given by
\begin{equation}
	e_L(\lambda,z) = \chi(\lambda) \exp(\pi (H - B)(z,\lambda) + \frac{\pi}{2}(H-B)(\lambda,\lambda)),
\end{equation}
for all $(\lambda,z) \in \Lambda \times \mathbb{R}^{2n}$, where we have defined $B$ to be the $\mathbb{C}$-bilinear extension of the symmetric form $\left.H\right|_{V_2 \times V_2}$. In order to bring these expressions into an explicit form, we note that there always exists a basis of $\mathbb{R}^{2n}$ such that $H$ can be written in the form \cite{Lange}
\begin{equation} \label{eq:H}
	H(v,w) = v^t (\textrm{Im} \Omega)^{-1} \bar{w},
\end{equation} 
where $\Omega$ satisfies $\Omega^t = \Omega$ and $\textrm{Im}\Omega >0$. In this basis $B$ takes the form 
\begin{equation} \label{eq:B}
	B(v,w) = v^t (\textrm{Im}\Omega)^{-1} w.
\end{equation}
Then $\Lambda = \Lambda_1 \oplus \Lambda_2$ with $\Lambda_1= \Omega \mathbb{Z}^n$ and $\Lambda_2 = D \mathbb{Z}^n$ is a decomposition for $H$. It induces a decomposition $\mathbb{C}^n = V_1 \oplus V_2$ with real vector spaces $V_1 = \Omega \mathbb{R}^n$ and $V_2 = \mathbb{R}^n$, and we can write every $z \in \mathbb{C}^n$ uniquely as 
\begin{equation}
	z = \Omega z^1 + D z^2,
\end{equation}
with $z^1, z^2 \in \mathbb{R}^n$. Note that this choice of coordinates determines the symplectic basis discussed previously and in which $\omega$ takes the form (\ref{eq:omega}). Using this basis, we find an explicit expression for $\chi(\lambda)$ given by
\begin{equation}
	\chi(\lambda) = \exp(\pi i \omega(\Omega \lambda^1,\lambda^2) + 2\pi i \omega(c,\lambda)),
\end{equation}
which for a principal polarization just becomes
\begin{equation}
	\chi(\lambda) = \exp(\pi i \lambda_1^t \lambda_2 + 2\pi i (\lambda_2^t c_1 - \lambda_1^t c_2)).
\end{equation}
Using this representation for the semichacter and the explicit expressions for $H$ and $B$ given in (\ref{eq:H}) and (\ref{eq:B}), one finds that the corresponding factor $e_L$ for a principal polarization is then given by
\begin{equation} \label{eq:eL}
	e_L(\lambda,z) = \exp(2\pi i (\lambda_2^t c_1 - \lambda_1^t c_2) - \pi i \lambda_1^t \Omega \lambda_1 - 2\pi i z^t \lambda_1).
\end{equation}
Sections of the line bundle $L$ will then transform with the above factor of automorphy under lattice shifts of the torus $J$. One then readily checks that the following theta function has exactly the right transformation properties\footnote{The theta function introduced here differs from the canonical theta function $\Theta^L$ by a prefactor which is irrelevant for our discussion here, see \cite{Lange} for further details.}
\begin{equation}
	\RT{c_1}{c_2}{\Omega}{z} \equiv \sum_{k \in \mathbb{Z}^n} \exp \left(\pi i (k + c_1)^t (\Omega(k+c_1) + 2(z+c_2))\right),
\end{equation}
i.e. it transforms with the factor $e_L$ given in (\ref{eq:eL}) and is thus an element of $H^0(L)$. 

Of course, for a principal polarization there is only one section as the space $H^0(L)$ is one-dimensional. But what happens for non-principal line bunldes? Now suppose $\mathcal{L} = (H,\chi)$ is a positive definite line bundle on $J$ and $c$ be a characteristic with respect to a decomposition $V = V_1 \oplus V_2$. For this case it is useful to define $K(\mathcal{L}) \equiv \Lambda(\mathcal{L})/\Lambda$.  Then the set $\{\Theta^\mathcal{L}_{\bar{w}} | \bar{w} \in K(\mathcal{L})_1\}$ with $K(\mathcal{L})_1 = K(\mathcal{L}) \cap V_1$ is a basis of the vector space $H^0(\mathcal{L})$ of theta functions for $\mathcal{L}$. Let us see what this means for a particular example. Consider the line bundle $\mathcal{L} = L^N$ for some positive integer $N$ where we take $L$ to be our already familiar line bundle with principal polarization. Then $\omega$ takes the form
\begin{equation}
	\omega = \left(\begin{array}{cc}0 & N \mathbf{1}_n \\ - N \mathbf{1}_n & 0 \end{array}\right),
\end{equation}
and furthermore we have $K(L^N)_1 = \frac{1}{N}\mathbb{Z}^n/\mathbb{Z}^n$. Fix $\bar{w}$ to be a representative of $K(L^N)_1$. Then the above just says that the theta functions
\begin{equation}
	\Theta^{L^N}_{\bar{w}} \sim \RT{c^1 + \omega}{c^2}{N\Omega}{N z}
\end{equation}
are linearly independent and form a basis of $H^0(L^N)$. What happens if we shift instead with $w \in K(L^N)_2$, namely a shift in $c^2$? In this case one can check that the theta functions merely acquire a constant factor
\begin{equation} \label{eq:c2shift}
	\RT{c^1}{c^2 + w}{N\Omega}{Nz} = e^{2\pi i \omega(c^1,w)} \RT{c^1}{c^2}{N\Omega}{Nz}.
\end{equation}
The current example was an example of a type $(N,N,\ldots,N)$ polarization. In Section \ref{sec:M5A} we will see an example of a type $(M,N)$ polarization and postpone the discussion of these more non-trivial line bundles to that Section.

\subsubsection*{Modular Transformation}

Apart from the shift-symmetry (\ref{eq:shiftsym}) there is another symmetry group under which theta functions transform covariantly, namely the group of symplectic $2n \times 2n$ integral matrices. For a given polarization $D$, this group is denoted by $Sp^D_{2n}(\mathbb{Z})$ and is defined as follows
\begin{equation}
	Sp^D_{2n} \equiv \left\{R = \left(\begin{array}{cc}\alpha & \beta \\ \gamma & \delta\end{array}\right) \in M_{2n}(\mathbb{Z}) \left| R \left(\begin{array}{cc}0 & D \\ -D & 0\end{array}\right) R^t = \left(\begin{array}{cc}0 & D \\ -D & 0\end{array}\right)\right.\right\}.
\end{equation}
Its action on $(\Omega,z)$ of a theta function is then given by
\begin{equation} \label{eq:Spaction}
	\left(\Omega,z\right) \mapsto \left((\alpha \Omega + \beta D)(D^{-1} \gamma \Omega + D^{-1} \delta D)^{-1},z^t (D^{-1} \gamma \Omega + D^{-1} \delta D)^{-1}\right).
\end{equation}
The exact transformation formula for a line bundle of arbitrary type can be found in \cite{Lange}. Here we want to focus on the case $\Theta^{L^N}_{\bar{w}}$ where $L=(H,\chi_0)$ is a principal line bundle with characteristic $c=0$. Here, $\bar{w}$ takes the values $\frac{a}{N}$, where $a \in (\mathbb{Z}/N)^n$. To this end, note that the group $Sp(2n,\mathbb{Z})$ is generated by matrices of the form \cite{Runge}
\begin{equation}
	T_S \equiv \left(\begin{array}{cc}\mathbf{1}_n & S \\ 0 & \mathbf{1}_n\end{array}\right),
\end{equation}
where $S$ runs over the symmetric $n\times n$-matrices, and the Fourier-transformation matrix 
\begin{equation} \label{eq:F}
	F \equiv \left(\begin{array}{cc}0 & \mathbf{1}_n \\ -\mathbf{1}_n & 0\end{array}\right).
\end{equation}
These generators then act as follows on the basis of theta functions discussed in the previous subsection \cite{Runge, Manschot:2008zb}
\begin{equation}
	T_s \RT{\frac{a}{N}}{0}{N \Omega}{N z} = \RT{\frac{a}{N}}{0}{N(\Omega+S)}{z} = e^{\pi i \frac{a^t S a}{N}}\RT{\frac{a}{N}}{0}{N \Omega}{N z},
\end{equation}
and
\begin{equation} \label{eq:SPtransform}
	F \RT{\frac{a}{N}}{0}{N\Omega}{N z} = \RT{\frac{a}{N}}{0}{N\Omega^{-1}}{N z\Omega^{-1}} =  \kappa \sum_{b \in (\mathbb{Z}/N)^n} (F_n)_{a,b} \RT{\frac{b}{N}}{0}{N \Omega}{N z},
\end{equation}
where
\begin{equation}
	\kappa = \sqrt{\textrm{det}-\Omega}~e^{\pi i N z^t {\Omega}^{-1}z}, \quad F_n = e^{2\pi i \frac{n}{8}} \left(\frac{1}{\sqrt{N}}\right)^n \left(\exp\left(2\pi i \frac{\langle a,b \rangle}{N}\right)\right)_{a,b \in (\mathbb{Z}/N)^n}.
\end{equation}
In the above $\langle a,b \rangle$ denotes the scalar product between the vectors $a$ and $b$. We notice that $\exp\left(2\pi i \frac{\langle a,b \rangle}{N}\right)$ is the perfect pairing which already appeared in the discussion of the S-duality transformation of the M5 brane wave functions in (\ref{eq:S-duality}). We will have more to say on this in Section \ref{sec:M5A} but to get there we need one more ingredient to which we turn now.

\subsection{Riemann's Addition Formula}
\label{sec:addition}

As we discussed in Sections \ref{sec:LST} and \ref{sec:Qstates}, theta functions can be viewed as states in a  Hilbert space corresponding to M5 branes on a certain background. Utilizing the operator-state correspondence these same theta functions can also be viewed as operators and one might then ask what is the operator product expansion for the product of two such operators? More precisely, we would like to view the M5 brane wave-functions as defect operators in a two-dimensional theory by taking into account the circular direction of the LST\footnote{Note that one of the dimensions corresponds to time and is already present in the quantum mechanical picture discussed so far}. In this picture, the fusion of two defect operators by moving the M5 branes on top of each other corresponds to evaluating the operator product of the individual defect operators. It turns out that these operators form a chiral ring and the corresponding product is Riemann's bilinear addition formula which relates products of sections of line bundles $L$ and $L'$ which have the same first Chern class $\omega$ to a sum of sections of $L \otimes L'$. This can schematically be written as follows:
\begin{equation} \label{eq:thetaproduct}
	\Theta^L_i(z) \cdot \Theta^{L'}_j(z) \sim \sum_{k \in H^0(L \otimes L')} c_{ijk} \Theta^{L \otimes L'}_k(z),
\end{equation}
where the $c_{ijk}$ are some constants not depending on the position on the Jacobian $J$. Before giving a precise version of this formula, let us reinterpret it in the context of M5 branes probing $S^1 \times \mathbb{C}^2/\Gamma$ where $\mathbb{C}^2/\Gamma$ is an A-type singularity (or also possibly a D-type singularity as we will see later). In this case theta functions for a line bundle $L$ of type $(M,N)$ correspond to states of the theory $\mathcal{T}_{M,N}$. Therefore, we see from (\ref{eq:thetaproduct}) that there must exist an operation which combines two copies of the theory $\mathcal{T}_{M,N}$ to a third as follows
\begin{equation} \label{eq:M5fusion}
	\mathcal{T}_{M,N} \otimes \mathcal{T}_{M,N} \longrightarrow \mathcal{T}_{2M,2N}.
\end{equation}
For M5 branes probing $S^1 \times \mathbb{C}^2$ it is clear what this means: we can combine one M5 brane with another parallel M5 brane to form the $A_1$ $(2,0)$ theory of two M5 branes, where we have restricted ourselves to the case of $M=1$ in (\ref{eq:M5fusion}). At the level of the Hilbert space this means that for two M5 branes probing a $\mathbb{Z}_{2N}$-singularity something nontrivial happens: the corresponding defect operator can in some cases be obtained by fusing the defect operators of single M5 branes probing $\mathbb{Z}_N$ singularities. The reason is that in the presence of the orbifold singularity we have fractional copies of the M5 branes which transform into each other under $\mathbb{Z}_{2N}$. Now since $\mathbb{Z}_N$ is a subgroup of $\mathbb{Z}_{2N}$, we can have two copies of states which transform into each other under $\mathbb{Z}_N$. These are the two copies corresponding to the two defect operators. This becomes clearer when we go to the T-dual frame where these $2N$ fractional copies will be actual M5 branes sitting at a point on the LST circle. These can then be separated into 2 copies of $N$ M5 branes each together with a $\mathbb{Z}_2$ symmetry interchanging them.

Let us now give a mathematically precise version of the formula (\ref{eq:thetaproduct}). To this end define a map $\alpha : J \times J \rightarrow J \times J$ by the map $(i,j) \mapsto (i+j,i-j)$. Then for all $(i,j) \in K(L) \times K(L')$ we have the following
\begin{eqnarray} \label{eq:productformula}
	~ & ~ & \RT{c^1_1 + i}{c_1^2}{\Omega}{z} \cdot \RT{c^1_2 + j}{c^2_2}{\Omega}{z} \nonumber \\
	~ & = & \sum_{k \in \half \mathbb{Z}_2} \RT{\half (c_1^1 - c_2^1) + k + \half(i-j)}{\half(c_1^2-c_2^2)}{2\Omega}{0} \RT{\half(c^1_1 + c^1_2) + k + \half(i+j)}{\half(c^2_1 + c^2_2)}{2\Omega}{2z}. \nonumber \\
\end{eqnarray}
As it turns out this is not the only relation between products of theta functions and one can show \cite{Lange} that there are also so-called \textit{cubic theta relations}:
\begin{equation} \label{eq:cubic}
	\Theta^{\mathcal{L}}_i \cdot \Theta^{\mathcal{L}}_j \cdot \Theta^{\mathcal{L}}_k \sim \sum_l c_{ijkl} \Theta^{\mathcal{L}^3}_l,
\end{equation}
where this time $\mathcal{L}$ is itself a cubic power of an ample line bundle $L$, i.e. $\mathcal{L} = L^3$. Together, the equations (\ref{eq:thetaproduct}) and (\ref{eq:cubic}) generate all non-trivial relations which theta functions of same first Chern class satisfy at the non-linear level. This allows one to embed abelian surfaces (or more generally abelian varieties) into projective space. To this end, one defines homogeneous coordinates as sections of line bundles on abelian surfaces as follows;
\begin{equation}
	X_{i,j} \equiv \RT{c^1 + (\frac{i}{M},\frac{j}{N})}{c^2}{\Omega}{z}, \quad \textrm{ where } \quad i=0,\ldots,M-1, \textrm{ and } j=0,\ldots, N-1.
\end{equation}
Then the abelian surface is defined through the projective space generated by these coordintes up to rescaling by $\lambda \in \mathbb{C}^*$ and the constrains generated by the theta relations:
\begin{equation} \label{eq:torusconstruction}
	\mathbb{T}\left[X_{i,j}\right] \equiv \frac{\mathbb{P}\left[X_{i,j}\right]}{\left[\Theta^2 \sim \Theta,~\Theta^3 \sim \Theta\right]}~.
\end{equation}
Although this might look a bit unfamiliar, such constructions are well-known to physicists in the case of the elliptic curve (i.e. a one-dimensional abelian variety). Here one defines
\begin{equation}
	X_i \equiv \theta[i/3,0](3\tau,3z), \quad i=0,1,2.
\end{equation}
Then the abstract construction (\ref{eq:torusconstruction}) boils down to \cite{Zaslow:2005wf}
\begin{equation} \label{eq:ellipticcurve}
	\mathbb{T}\left[X_i\right] \equiv \frac{\mathbb{P}^2\left[X_i\right]}{\left[X_0^3 + X_1^3 + X_1^3 + \mu X_0 X_1 X_2 = 0\right]},
\end{equation}
which is the familiar construction of the elliptic curve through an algebraic constraint in $\mathbb{P}^2$. Equation (\ref{eq:torusconstruction}) is just a version of (\ref{eq:ellipticcurve}) in one dimension higher. In this case, however, the number of constraints is vastly higher than in the case of the elliptic curve and we refer to \cite{Keiichi} for further details. Now it is time to apply our theta function technology to M5 branes to which we turn next.

\subsection{M5 branes probing A-type singularities}
\label{sec:M5A}

Let us consider $M$ M5 branes probing $S^1 \times \mathbb{C}^2/\mathbb{Z}_N$  in M-theory. As discussed in Sections \ref{sec:LST} and \ref{sec:Qstates}, the quantum states of this theory correspond to theta functions which are sections of a line bundle $L$ of type $(M,N)$ on an abelian surface. In fact, such sections can be described rather explicitly through theta functions
\begin{equation}
	\RT{(\frac{i}{M},\frac{j}{N})}{0}{\left[\begin{array}{cc}M \tau & \sigma \\ \sigma & N \rho \end{array}\right]}{ \left(\begin{array}{c}M z_1 \\ N z_2 \end{array}\right)}, \quad i=0,\ldots,M-1 \textrm{ and } j=0, \ldots,N-1.
\end{equation}
Here we have focused on a line bundle with characteristic function $\chi_0$ such that the values of $c^1$ and $c^2$ are zero modulo integers to keep things simple and note that the analysis which we are going to perform in this Section can be carried out for arbitrary characteristic. For convenience we will henceforth define
\begin{equation}
	\Omega \equiv \left(\begin{array}{cc}M \tau & \sigma \\ \sigma & N \rho \end{array}\right), \quad \widetilde{\Omega} \equiv \left(\begin{array}{cc}\tau & \sigma/M \\ \sigma/N & \rho \end{array}\right), \quad \textrm{ such that } \left(\begin{array}{cc}M & 0 \\ 0 & N\end{array}\right) \widetilde{\Omega} = \Omega.
\end{equation}
Note, however, that the $Sp^D_{4}(\mathbb{Z})$ transformation given in (\ref{eq:Spaction}) is still defined as an action on $\Omega$. For instance, in the case of $M = N$ giving $\Omega = N \widetilde{\Omega}$, the element $F$ (see (\ref{eq:F})) acts as follows
\begin{equation}
	F(\Omega) = (0 \cdot \Omega + N)(-N^{-1} \Omega + 0)^{-1} = -N {\widetilde{\Omega}}^{-1},
\end{equation}
reproducing the transformation (\ref{eq:SPtransform}).
Then we can compute the transformation behavior of our theta functions under shifts $\vec{z} \mapsto \vec{z} + \widetilde{\Omega} \vec{\lambda} + \vec{\mu}$ (here $\vec{\lambda}, \vec{\mu} \in \mathbb{Z}^2$)\footnote{We define $\exp\left[\cdot\right] \equiv e^{2\pi i \cdot}$}:
\begin{eqnarray}
	~ & ~ & \RT{(\frac{i}{M},\frac{j}{N})}{0}{\Omega}{\left(\begin{array}{cc}M & 0 \\ 0 & N\end{array}\right) \cdot \left(\vec{z} + \widetilde{\Omega}\vec{\lambda} + \vec{\mu}\right)} \nonumber \\
	~ & = & \sum_{\vec{n} \in \mathbb{Z}^2} \exp\left[\half \vec{n'}^t \Omega \vec{n'} + \vec{n'} \cdot \left(\left(\begin{array}{cc}M & 0 \\ 0 & N\end{array}\right) \cdot \left(\vec{z} + \widetilde{\Omega}\vec{\lambda} + \vec{\mu}\right)\right)\right], \quad \textrm{ where } \quad \vec{n'} = \vec{n} + \left(\begin{array}{c}\frac{i}{M} \\ \frac{j}{N}\end{array}\right) \nonumber \\
	~ & = & \sum_{\vec{n} \in \mathbb{Z}^2} \exp\left[\half \left(\vec{n'} + \vec{\lambda}\right)^t \Omega \left(\vec{n'} + \vec{\lambda}\right) + \left(\vec{n'}+ \vec{\lambda}\right)^t  \left(\begin{array}{cc}M & 0 \\ 0 & N\end{array}\right)\vec{z}\right. \nonumber \\
	~ & ~ & \left.  - \half \vec{\lambda}^t \Omega \vec{\lambda} + \vec{n'}^t \left(\begin{array}{cc}M & 0 \\ 0 & N\end{array}\right) \vec{\mu} - \vec{\lambda}^t \left(\begin{array}{cc}M & 0 \\ 0 & N\end{array}\right) \vec{z} \right] \nonumber \\
	~ & = & e_L(\vec{\lambda},\vec{\mu},\vec{z}) \RT{(\frac{i}{M},\frac{j}{N})}{0}{\Omega}{\left(\begin{array}{cc}M & 0 \\ 0 & N\end{array}\right)\vec{z}}, \nonumber \\
\end{eqnarray}
with
\begin{equation}
	e_L(\vec{\lambda},\vec{\mu},\vec{z}) = \chi(\vec{\lambda},\vec{\mu}) \exp\left[-\half \vec{\lambda}^t \Omega \vec{\lambda} - \vec{\lambda}^t \left(\begin{array}{cc}M & 0 \\ 0 & N\end{array}\right) \vec{z} \right],
\end{equation}
and
\begin{equation} \label{eq:chi}
	\chi(\vec{\lambda},\vec{\mu}) = \exp\left[\vec{c^1}^t \left(\begin{array}{cc}M & 0 \\ 0 & N\end{array}\right)\vec{\mu} - \vec{c^2}^t \left(\begin{array}{cc}M & 0 \\ 0 & N\end{array}\right) \vec{\lambda}\right].
\end{equation}
In the last line we have restored the dependence on the characteristics $c^1$ and $c^2$. For our particular case $\vec{c^1}^t = \left(\frac{i}{M},\frac{j}{N}\right)$ and $\vec{c^2}^t = 0$. Thus we see that $\chi(\vec{\lambda},\vec{\mu}) = 1$ for all $i=0, \ldots, M-1$ and $j = 0, \ldots, N-1$. Therefore, our theta functions are all sections of the same line bundle, i.e.
\begin{equation}
	\RT{(\frac{i}{M},\frac{j}{N})}{0}{\Omega}{\left(\begin{array}{c}M z_1 \\ N z_2 \end{array}\right)} \in H^0(L).
\end{equation}
From (\ref{eq:chi}) we also see that the first Chern class of our line bundle $L$ is given by
\begin{equation}
	\omega_L = \left(\begin{array}{cc} 0 & D \\ -D & 0\end{array}\right), \quad \textrm{ where } D = \left(\begin{array}{cc}M & 0 \\ 0 & N\end{array}\right).
\end{equation}
Thus we have shown that our line bundle is of polarization $(M,N)$. From our construction we can also immediately see the quotient description of our abelian surface $\mathbb{T}_{\Omega}$. As our theta functions transform covariantly under shifts by $\widetilde{\Omega}\vec{\lambda}$ and $\vec{\mu}$, we can use exponentiated coordinates 
\begin{equation}
	X \equiv e^{2\pi i z_1} \quad \textrm{ and } \quad Y \equiv e^{2\pi i z_2},
\end{equation} 
such that $\mathbb{T}_{\Omega}$ is the quotient $(\mathbb{C}^*)^2/\mathbb{Z}^2$, where the generators of $\mathbb{Z}^2$ act by
\begin{equation}
	(X,Y) \mapsto (e^{2\pi i \tau} X, e^{2\pi i \sigma/N} Y), \quad (X,Y) \mapsto (e^{2\pi i \sigma/M} X, e^{2\pi i \rho} Y).
\end{equation}
Equivalently, our surface is given by the construction (\ref{eq:torusconstruction}) with homogeneous coordinates $X_{i,j}$ given by our theta functions, i.e.
\begin{equation}
	X_{i,j} \equiv \RT{(\frac{i}{M},\frac{j}{N})}{0}{\Omega}{\left(\begin{array}{c}M z_1 \\ N z_2 \end{array}\right)} \quad \textrm{ and } \quad\mathbb{T}_{\Omega} = \mathbb{T}\left[X_{i,j}\right].
\end{equation}
It turns out that this second construction is more useful for our purposes as it immediately also allows us to write down the moduli space of vacua for our theory and the corresponding Seiberg-Witten curve.

\subsubsection*{Moduli space of Vacua}

Following the discussion in \ref{sec:Qstates} the Seiberg-Witten curve for our LST is given by the hypersurface
\begin{equation} \label{eq:ASWcurve}
	\mathcal{W}(X_{i,j})  \equiv \sum_{i,j} a_{i,j} X_{i,j} = 0
\end{equation}
in $\mathbb{T}[X_{i,j}]$. This is the equation of a Riemann surface in our abelian surface which we henceforth shall denote by $\Sigma$. This is exactly the same Seiberg-Witten curve already obtained using different methods in \cite{Haghighat:2017vch} (see also \cite{Kanazawa:2016tnt}). Viewing the $X_{i,j}$ as operators as discussed previously, we can also identify $\Sigma$ with the corresponding chiral ring as follows
\begin{equation} \label{eq:LGcurve}
	\Sigma = \frac{\mathbb{T}[X_{i,j}]}{\left[\mathcal{W}(X_{i,j})\right]}.
\end{equation}
Let us compute the genus of $\Sigma$. Assuming that our line bundle $L$ is ample, the Riemann-Roch Theorem implies
\begin{equation}
	h^0(L) = \half (L^2) = M N.
\end{equation}
From the adjunction formula we know that $\left. T J\right|_{\Sigma} = N \Sigma \oplus T \Sigma$ where the symbols $T$ and $N$ denote tangent and normal bundles, respectively. Thus integrating the first Chern classes over $\Sigma$ gives
\begin{equation}
	2 g_{\Sigma} -2 = (\Sigma^2) = 2 M N,
\end{equation}
and from this we learn that $g_{\Sigma} = M N + 1$. The moduli space of the Seiberg-Witten curve is then given by the collection of parameters $a_{i,j}$ up to rescaling by $\lambda \in \mathbb{C}^*$ which in turn gives the space $\mathbb{P}^{MN-1}$. However, we have to bear in mind that elements of $Sp^D_4(\mathbb{Z})$ act non-trivially on the $X_{i,j}$, i.e. we have for any $G \in Sp^D_4(\mathbb{Z})$
\begin{equation}
	X_{i,j}(G(\Omega),G(z)) = \sum_{i',j'} \Gamma(G)_{i,j,i',j'} X_{i',j'}.
\end{equation}
An important property of the $\Gamma(G)$ is that they furnish an irreducible representation of $Sp^D_4(\mathbb{Z})$ and hence are invertible \cite{Lange}\footnote{We would like to thank Gerard van der Geer for clarifying this point in a private communication.}. Therefore, for the equation (\ref{eq:ASWcurve}) to stay invariant, we need the parameters $a_{i,j}$ to transform as follows
\begin{equation} \label{eq:SPaction}
	a_{i,j} \mapsto \sum_{i',j'} (\Gamma^{-1})_{i,j,i',j'}~a_{i',j'}.
\end{equation}
This shows that the total moduli space is a fibration
\begin{equation}
	\mathcal{M}_{M,N} \equiv \left(\mathbb{C}^{MN} \times \mathbb{H}_2\right)/\left(Sp(4,\mathbb{Z}) \times \mathbb{C}^*\right),
\end{equation}
where $\mathbb{H}_2 = \{\Omega \in M_2(\mathbb{C}) | \Omega^t = \Omega, \textrm{Im}\Omega > 0\}$ is the Siegel upper half space and the action of $Sp^D_4(\mathbb{Z})$ on $\mathbb{C}^{MN}$ is given by (\ref{eq:SPaction}) and on $\mathbb{H}_2$ by (\ref{eq:SPtransform}). As $\mathbb{H}_2$ is complex three-dimensional, this shows that the total moduli space is of dimension
\begin{equation}
	\textrm{dim}_{\mathbb{C}} \mathcal{M}_{M,N} = M N + 2.
\end{equation}
Of course, it can happen that for certain choices of $\Omega$, i.e. at certain points on the moduli space, we obtain fix points under the action of a subgroup of $Sp^D_4(\mathbb{Z})$. Then at such points, $\Gamma(G)$ would act as the identity for elements $G$ of that particular subgroup. But this would still lead to a well-defined equation (\ref{eq:ASWcurve}) as the corresponding action on the $a_{i,j}$ can be taken to be the identity as well. If for some reason, in order to construct a smooth fibration for example, one needs a freely acting subgroup of $Sp^D_4(\mathbb{Z})$, one can always resort to 
\begin{equation}
	Sp^D_4(D) \equiv \left\{\left(\begin{array}{cc}\alpha & \beta \\ \gamma & \delta\end{array}\right) \in Sp^D_4(\mathbb{Z}) \left| \alpha - \mathbf{1}_2 = b = c = d-\mathbf{1}_2 = 0 ~\mod D\right.\right\}.
\end{equation}

\subsubsection*{Duality web}

So far we have not put any specific constraints on the integer numbers $M$ and $N$. More specifically, one can write
\begin{equation}
	M = p M', \quad N = p N', \quad \textrm{ where } p = \textrm{gcd}(M,N).
\end{equation}
Then $M'$ and $N'$ are co-prime and $p$ is some integer number. To understand the role of $p$ let us consider a specific example, namely $p=2$. Then the first Chern class of our line bundle $L$ can be written as twice the first Chern class of another line bundle $L'$
\begin{equation}
	\omega_L = 2 \omega_{L'},
\end{equation}
such that $L'$ is a line bundle of polarization $(M',N')$. In this case we will have $L \cong L'^2$ and we can view our theory as a gluing of two theories, namely
\begin{equation}
	\mathcal{T}_{M',N'} \otimes \mathcal{T}_{M',N'} \rightarrow \mathcal{T}_{M,N}.
\end{equation}
Note that we cannot repeat this process as $M'$ and $N'$ are now co-prime. This means that there must be a duality frame where $\mathcal{T}_{M',N'}$ corresponds to a single M5 brane probing some singularity and the above then corresponds to the gluing of two such M5 brane theories. We can also see this at the level of the Seiberg-Witten curve where $\mathcal{W}$ can now (for some choice of moduli $a_{i,j}$) be written as 
\begin{equation}
	\mathcal{W}(X_{i,j}) = \mathcal{W}_1(X_{i,j}') \cdot \mathcal{W}_2(X_{i,j}'),
\end{equation}
where the $X_{i,j}'$ are now sections of $L'$. This means that our Seiberg-Witten curve has degenerated to a reducible configuration consisting of two curves of genus $M' N' + 1$ whose equation is given by 
\begin{equation}
	\mathcal{W}_1 = 0 \quad \textrm{ or } \quad \mathcal{W}_2 = 0.
\end{equation}
Thus the moduli space has a limit where it splits into two components corresponding to the moduli spaces of the two seperated M5 branes. This is a specific instance of the general statement observed in \cite{Hohenegger:2016yuv} where we are now in the mirror dual frame: LSTs of type $(M,N)$ are equivalent to LSTs of type $(p, MN/p)$. We can now give a proof of this statement using simple properties of theta functions already discussed. Recall that our abelian surface is given as the quotient $J = \mathbb{R}^{2n}/\Lambda$. We also know that our line bundle $L$ of polarization $(M,N)$ has first Chern class
\begin{equation}
	\omega = \left(\begin{array}{cc}0 & D \\ - D & 0\end{array}\right), \quad D = \left(\begin{array}{cc}M & 0 \\ 0 & N\end{array}\right).
\end{equation} 
Thus from our point of view, all we have to do to show the duality is to find a unimodular basis transformation, i.e. $SL(4,\mathbb{Z})$ transformation, which keeps the lattice $\Lambda$ invariant and changes the anti-symmetric form $\omega$ to
\begin{equation} \label{eq:normalform}
	\omega_L \mapsto \left(\begin{array}{cc}0 & \widetilde{D} \\ - \widetilde{D} & 0\end{array}\right), \quad \widetilde{D} = \left(\begin{array}{cc}p & 0 \\ 0 & M N/p\end{array}\right).
\end{equation}
This can be easily achieved by the following congruence transformation
\begin{equation}
	\left(\begin{array}{cc}P^t & 0 \\ 0 & Q^t\end{array}\right) \left(\begin{array}{cc}0 & D \\ -D & 0\end{array}\right) \left(\begin{array}{cc}P & 0 \\ 0 & Q\end{array}\right) = \left(\begin{array}{cc}0 & P^t D Q \\ - Q^t D P & 0\end{array}\right),
\end{equation}
with $P$ and $Q$ $SL(2,\mathbb{Z})$ matrices chosen such that they transform the matrix $D$ into Smith normal form. For example, the smith decomposition for $M=8, N=22$ is 
\begin{equation}
	\left(\begin{array}{cc}-8 & 3 \\ -11 & 4\end{array}\right)\left(\begin{array}{cc}8 & 0 \\ 0 & 22\end{array}\right) \left(\begin{array}{cc}1 & -33\\ 1 & -32\end{array}\right) = \left(\begin{array}{cc}2 & 0 \\ 0 & 88\end{array}\right).
\end{equation}
The fact that a unimodular transformation can be always found which brings $\omega$ into the form (\ref{eq:normalform}) is a Theorem, namely Theorem 6.1 of reference \cite{Kaplansky}. This concludes our proof of the duality observed in \cite{Hohenegger:2016yuv} in the sense that we can explain its origin in the freedom to choose a polarization for the underlying quantum wave functions. In our picture the partition fucntions studied in \cite{Hohenegger:2016yuv} appear as multiplicative factors of the wave-functions studied in our paper and capture the dependence on the background metric including $\epsilon_1$ and $\epsilon_2$. A generic partition function can then be seen as a state in a Hilbert space $\Psi(\Omega) \in \mathcal{H}$ which can be expanded in the basis of theta-functions we have found as follows \cite{Witten:1998wy}
\begin{equation}
	\Psi(\Omega) = \sum_{w} Z_w(\Omega) \Psi_w, \quad w \in \mathbb{Z}_N \times \mathbb{Z}_M.
\end{equation} 
We would like to point out that for the particular case $\textrm{gcd}(N,M) = p = 1$, a proof of the duality at the level of the $Z_w(\Omega)$ (for all values of $\epsilon_{1,2}$) was also given in \cite{Bastian:2017ing}.
Note that for $p > 1$ the corresponding M5 brane wave-functions transform into each other under S-duality according to (\ref{eq:SPtransform}) with $N=p$. One important difference to the analysis of \cite{Bastian:2017ing} is that in our case the 6d theory is living on $T^2 \times \widehat{\mathbb{C}^2}$ where as in \cite{Bastian:2017ing} the authors consider a spacetime of the form $T^2 \times_{\epsilon_1,\epsilon_2} \mathbb{C}^2$ which has trivial intermediate Jacobian and hence the Hilbert space is one-dimensional in their case.

\subsection{M5 branes probing D-type singularities}
\label{sec:M5D}

\begin{figure}[h]
  \centering
	\includegraphics[width=\textwidth]{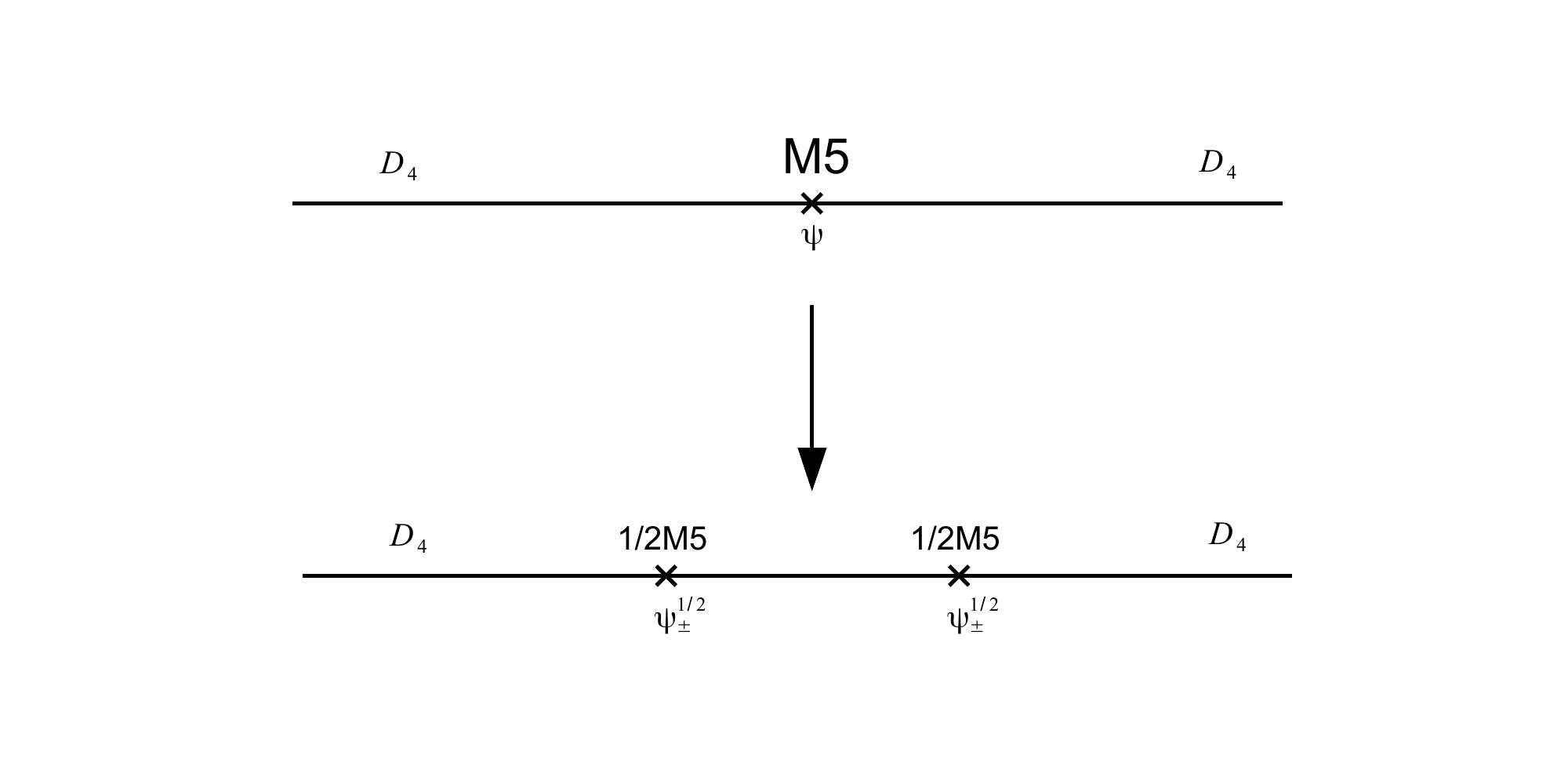}
  \caption{M5 branes probing D-type singularities.}
  \label{fig:M5D}
\end{figure}

In this Section we want to construct wave functions for M5 branes probing $D$-type singularities. Here our discussion will be enriched by a fundamental new symmetry, namely the $\mathbb{Z}_2$ reflection symmetry of the root lattice of $D_n$ as already observed in \cite{Haghighat:2018dwe}, which acts on our wave functions and under which they have to be invariant. Consider in the following the LST arising from a single M5 brane probing a singularity of type $D_4$, i.e. the perpendicular space to the M5 brane is $S^1 \times \mathbb{C}^2/\Gamma_{D_4}$. We shall denote this theory by $\mathcal{T}_{1,D_4}$. In this case there is a further effect which was observed in \cite{DelZotto:2014hpa}, namely the M5 brane splits into two fractional $\half \textrm{M5}$ branes. Then our $\mathbb{Z}_2$ symmetry will act now on these fractional branes and their wavefunctions will be either odd or even under this action. The reason is that only their square which was our original M5 brane has to be invariant. Let us denote the $\half\textrm{M5}$ wavefunctions by $|\psi^{\half}_{\pm}\rangle$ to keep track of the $\mathbb{Z}_2$ Eigenvalues. This is schematically depicted in Figure \ref{fig:M5D}.

Let us now consider the T-dual picture of our LST where we first compactify on one of the internal directions of our M5 brane and then perform T-duality along the perpendicular $S^1$ anoalogous to the already discussed $A$-type case. Performing these steps we arrive at the $D_4$ $(2,0)$ theory probing $S^1\times TN_1$, where $TN_1$ is one-centered Taub-Nut. The $D_4$ $(2,0)$ theory can be equivalently viewed as $4$ $\textrm{M5}$ branes sitting on top of an M-theoretic orientifold plane, see \cite{deBoer:1998by,Hori:1998iv,Tachikawa:2009rb} for further details. This second picture is particularly useful for us as it allows us to ``see" the action of our $\mathbb{Z}_2$ symmetry explicitly in this T-dual frame. Here it is nothing else than the orientifold action. One can think of these $4$ $\textrm{M5}$ branes as sitting on one side of the orientifold plane with $4$ mirror images sitting on the other side. Then one can find exactly $4$ linear combinations of our $\textrm{M5}$ branes and their mirrors which are either odd or even under the $\mathbb{Z}_2$ action. In fact, as we will see later, we will have exactly $3$ combinations which are even and one linear combination which is odd. Wavefunctions of the full theory are then squares of these wavefunctions. 

Putting the two duality frames together we again see that our wavefunctions must be sections of line bundles on $\mathbb{T}_{\Omega} = \mathbb{T}_{\tau} \times \mathbb{T}_{\rho}$ where this time the off-diagonal components of $\Omega$ are zero as there is no mass-deformation in the D-type theory. The action of $\mathbb{Z}_2$ is then nothing else than the reflection \cite{Haghighat:2018dwe}
\begin{equation} \label{eq:reflection}
	(-1)^*_{\mathbb{T}_{\Omega}} : \quad z \mapsto - z , \quad \textrm{for} \quad z \in \mathbb{T}_{\Omega}.
\end{equation}
From the above discussion we also see that states of the theory $\mathcal{T}_{1,D_4}$ must be sections of a line bundle $L$ of polarization $(2,8)$ as we have $2$ $\half\textrm{M5}$ branes in one duality frame and $8$ $\half\textrm{M5}$ branes in the other. Moreover, these sections must be products of sections of type $(1,4)$. In order to construct these sections we first have to determine how the operator $(-1)^*_{\mathbb{T}_{\Omega}}$ acts on theta functions.

\subsubsection*{Even and odd Theta Functions}
As outlined above, in order to proceed we need to find out how the reflection (\ref{eq:reflection}) acts on the space of sections of line bundles on $\mathbb{T}_{\Omega}$, namely our theta functions. Here we want to describe this action explicitly. To this end, suppose now that $L = (H,\chi)$ is an ample line bundle on $J$. One can easily convince oneself that for the action (\ref{eq:reflection}) to be a symmetry of $L$, i.e. $L$ remains invariant such that $(-1)^*_J L \cong L$, the semicharacter $\chi$ has to take the values $\pm 1$. This is only possible if $c \in \half \Lambda(L)$ and such line bundles are called \textit{symmetric}. For such line bundles the reflection (\ref{eq:reflection}) becomes an isomorphism and induces an involution on the vector space of theta functions for $L$
\begin{equation}
	(-1)^*_J : H^0(L) \rightarrow H^0(L), \quad \Theta^L(z) \mapsto \Theta^L(-z).
\end{equation}
Denote by $H^0(L)_+$ and $H^0(L)_-$ the eigenspaces of the involution $(-1)^*_J$. For the computation of the dimensions $h^0(L)_+$ and $h^0(L)_-$, we need to work out how $(-1)_J$ acts on $H^0(L)$. For this choose a decomposition $\Lambda = \Lambda_1 \oplus \Lambda_2$ for $L$ giving rise to a decomposition $c = c_1 + c_2$ of the characteristic $c \in \half \Lambda(L)$. Then we have
\begin{equation} \label{eq:inversion1}
	(-1)^*_J ~\RT{c_1+w}{c_2}{\Omega}{z} = \exp\left(4 \pi i \omega(w+c_1,c_2)\right) \RT{-c_1 - w}{c_2}{\Omega}{z}, \quad w \in K(L)_1.
\end{equation}
This formula simplifies considerably for characteristic zero where we obtain
\begin{equation} \label{eq:inversion2}
	(-1)^*_J~ \RT{w}{0}{\Omega}{z} = \RT{-w}{0}{\Omega}{z}, \quad \textrm{ for all } w \in K(L)_1.
\end{equation}
In this case it is also easy to write down eigenfunctions for $(-1)^*_J$. Simply define
\begin{equation}
	\Theta^{\pm}_w \equiv \RT{w}{0}{\Omega}{z} \pm \RT{-w}{0}{\Omega}{z},
\end{equation}
It follows immediately from (\ref{eq:inversion2}) that $\Theta^+_w$ is an even function and $\Theta^{-}_w$ is odd. Since $\{\Theta^+_w, \Theta^-_w |~ w \in K(L)_1\}$ spans the vector space $H^0(L)$, the theta functions $\Theta^+_w$, $w \in K(L)_1$ span $H^0(L)_+$. Using this basis one can show after some thought that for a line bundle of type $(d_1,\ldots,d_n)$, with characteristic $0$, we have
\begin{equation} \label{eq:dimpm}
	h^0(L_0)_{\pm} = \half h^0(L_0) \pm 2^{n-s-1},
\end{equation}
where the number $s$ is obtained from a decomposition $d_1,\ldots, d_s$ of odd numbers and $d_{s+1},\ldots,d_n$ of even numbers. 

\subsubsection*{Construction of the Hilbert space}
Let us now apply the above results to our picture of a single M5 brane probing a $D_4$ singularity. In the following we shall construct the space of ground states of the M5 brane. Following our previous heuristic discussion, we start in the first duality frame, construct the corresponding states and then turn to the second duality frame and repeat the procedure there. At the end we will put everything together. 

In the first duality frame, we have two $\half\textrm{M5}$ branes and their corresponding operators/states are given by the following two theta functions
\begin{eqnarray}
	\theta_1(\tau,z_1) & = & \RT{1/2}{1/2}{\tau}{z_1}, \nonumber \\
	\theta_4(\tau,z_1) & = & \RT{0}{1/2}{\tau}{z_1}.
\end{eqnarray}
Let us explain the appearance of these particular theta functions here. The second characteristic is for both $c_2 = \half$. This means that both states are twisted sector states under the $\mathbb{Z}_2$ action and transform with a minus sign (see equation (\ref{eq:eL})) under shifts along the periodic direction $\tau$. The first characteristic is different, namely $c_1 = \half$ in one case and $c_1 = 0$ in the other. Looking at (\ref{eq:inversion1}), this directly tells us that the first state is odd under the $\mathbb{Z}_2$ reflection while the second one is even. They thus correspond to our wavefunctions $\psi^{\half}_{\pm}$. Note that $\theta_1$ and $\theta_4$ are sections of \textit{different} line bundles over $\mathbb{T}_{\tau}$! However, $\mathbb{Z}_2$-invariant states appear only at the level $\left(\psi^{\half}_{\pm}\right)^2$ and indeed the squares of $\theta_1$ and $\theta_4$ belong to the same line bundle as one can readily check using (\ref{eq:productformula}):
\begin{eqnarray}
	\theta_1(\tau,z_1)^2 & = & \RT{1/2}{0}{2\tau}{0} \RT{0}{0}{2\tau}{2z_1} - \RT{0}{0}{2\tau}{0}\RT{1/2}{0}{2\tau}{2z_1}, \nonumber \\
	\theta_4(\tau,z_1)^2 & = & \RT{0}{0}{2\tau}{0}\RT{0}{0}{2\tau}{2z_1} - \RT{1/2}{0}{2\tau}{0}\RT{1/2}{0}{2\tau}{2z_1}. \nonumber \\
\end{eqnarray}

Let us now turn to the T-dual frame. Here, following our previous discussion, we want to construct sections of a degree $4$ line bundle over $\mathbb{T}_{\rho}$. Such a line bundle will have always $4$ linearly independent sections. A glance at formula (\ref{eq:dimpm}) tells us there must be exactly $3$ linear combinations which are even under the $\mathbb{Z}_2$ reflection and one combination which is odd. Let us see what these linear combinations are:
\begin{eqnarray}
	\Theta^+_1 & = & \RT{0}{0}{4\rho}{4z_2} + \RT{1/2}{0}{4\rho}{4z_2}, \nonumber \\
	\Theta^+_2 & = & \RT{1/4}{0}{4\rho}{4z_2} + \RT{3/4}{0}{4\rho}{4z_2}, \nonumber \\
	\Theta^+_{3} & = & \RT{0}{0}{4\rho}{4z_2} - \RT{1/2}{0}{4\rho}{4z_2}, \nonumber \\
	\Theta^-_{1} & = & \RT{1/4}{0}{4\rho}{4z_2} - \RT{3/4}{0}{4\rho}{4z_2}.
\end{eqnarray}
These linear combinations correspond to even and odd M5 brane states under the orientifold action. These states by themselves are not $\mathbb{Z}_2$ invariant and we have to multiply them with suitable mirror states to form invariant states. Such invariant states will be even sections of degree $8$ line bundles. To see this, one can use the addition formula (\ref{eq:productformula}) and compute for example
\begin{eqnarray}
	\Theta^+_1 \cdot \Theta_2^+ & = & \RT{1/2}{0}{4\rho}{4z_2} \RT{1/4}{0}{4\rho}{4z_2} + \ldots, \nonumber \\
	~ & = & \RT{1/8}{0}{8\rho}{0} \RT{3/8}{0}{8\rho}{8z_2} + \RT{5/8}{0}{8\rho}{0} \RT{7/8}{0}{8\rho}{8z_2} + \ldots \nonumber \\
\end{eqnarray}
Note that taking all possible products $\Theta^{+}_i \Theta^{+}_j$ and $\left(\Theta_1^{-}\right)^2$ gives all even sections of our degree $8$ line bundle as for such a line bundle we have $h^0(L^8)_+ = 5$\footnote{Here $L$ is understood to be principal.}.

Now it is time to put everything together. Here we will need the following essential theta function identity
\begin{equation} \label{eq:bigtheta}
	\RT{a_1}{b_1}{M\tau}{Mz_1} \RT{a_2}{b_2}{N\rho}{Nz_2} = \RT{(a_1,a_2)}{(b_1,b_2)}{\Omega}{(M z_1,N z_2)},
\end{equation}
where $\Omega = \left(\begin{array}{cc}M\tau & 0 \\ 0 & N\rho\end{array}\right)$. Then we can define
\begin{eqnarray}
	X_i & \equiv & \theta_1(\tau,z_1) \Theta^+_i(\rho,z_2), \nonumber \\
	Y    & \equiv & \theta_4(\tau,z_1) \Theta^-_1(\rho,z_2).
\end{eqnarray}
Note that $X_i$ and $Y$ can be fully expressed in terms of the ``big" theta function (\ref{eq:bigtheta}), for example we have
\begin{equation} \label{eq:bigthetaX}
	X_1 = \RT{(1/2,0)}{(1/2,0)}{\left(\begin{array}{cc}\tau & 0 \\ 0 & 4 \rho\end{array}\right)}{(z_1,4z_2)} + \RT{(1/2,1/2)}{(1/2,0)}{\left(\begin{array}{cc}\tau & 0 \\ 0 & 4 \rho\end{array}\right)}{(z_1,4z_2)}.
\end{equation}
Under our $\mathbb{Z}_2$ reflection we have
\begin{equation}
	X_i \mapsto - X_i, \quad  Y \mapsto -Y,
\end{equation}
which shows that none of these coordinates are invariant. This is as it should be because $\half\textrm{M5}$ brane states are not eigenstates of the reflection but their products are. So let us look at the operators which appear at second order:
\begin{equation}
	X_i^2,\quad  X_i X_j \textrm{ for } i < j, \quad Y^2.
\end{equation}
We don't have combinations $X_i Y$ as these are sections of a different line bundle as can be easily seen from the addition formula (\ref{eq:productformula}). The states which appear at second order as shown above, are exactly the states of our M5 brane probing the $D_4$ singularity. Thus the corresponding Seiberg-Witten curve parametrizing the moduli space of vacua is given by the hypersurface
\begin{equation} \label{eq:DtypeSP}
	\mathcal{W} = \sum_{i\leq j} a_{i,j} X_i X_j + b Y^2 = 0
\end{equation}
in $\left(\mathbb{T}_{\Omega} = \mathbb{T}[X_i,Y]\right)/\mathbb{Z}_2$. It is amusing to note that the number of independent terms in (\ref{eq:DtypeSP}) is $6$ which is the dual coxeter number of $D_4$. 

We end this Section by proving that the Seiberg-Witten curve given in (\ref{eq:DtypeSP}) matches exactly the result obtained in \cite{Haghighat:2018dwe} where different methods were used. To see this, note that the Seiberg-Witten curve there was given by
\begin{equation} \label{eq:SWcurve}
	0 = \frac{\theta_3(\tau,0)^2\theta_2(\tau,0)^2}{4 \eta(\tau)^6} \cdot \sum_{n=0}^4 a_n X(\rho,z_2)^n - \left(\frac{\theta_3(\tau,0)^4 + \theta_2(\tau,0)^4}{12} + \frac{X(\tau,z_1)}{4}\right) \cdot \frac{c_0}{64}Y(\rho,z_2)^2,
\end{equation}
where $X$ and $Y$ are the Weierstrass functions
\begin{eqnarray}
	X(\rho,z) & = & \theta_3(\rho,0)^2 \theta_2(\rho,0)^2 \frac{\theta_4(\rho,z)^2}{\theta_1(\rho,z)^2} - \frac{\theta_3(\rho,0)^4+\theta_2(\rho,0)^4}{3},  \nonumber \\
	Y(\rho,z)^2 & = & 4 X(\rho,z)^3 - \frac{4}{3} E_4(\rho) X(\rho,z) - \frac{8}{27} E_6(\rho),
\end{eqnarray}
with $E_4$, $E_6$ and $\eta$ the Eisenstein series of index $4$ and $6$ respectively and the Dedekind eta function. Multiplying (\ref{eq:SWcurve}) with $\theta_1(\tau,z_1)^2\theta_1(\rho,z_2)^8$ we see that the result can be expressed in the form
\begin{equation} \label{eq:SWreformulated}
	0 = \theta_1(\tau,z_1)^2 \left(\sum_{2i+2j=8} \Delta^{(1)}_{i,j} \theta_4(\rho,z_2)^{2i} \theta_1(\rho,z_2)^{2j}\right) + \theta_4(\tau,z_1)^2 \left(\Delta^{(2)} \theta_1(\rho,z_2)^8 Y(\rho,z_2)^2\right).
\end{equation}
Now we make the following crucial observation:
\begin{equation}
	\theta_1(\rho,z_2)^4 Y(\rho,z_2) = 4 \eta(\rho)^9 \Theta^-_1(\rho,z_2).
\end{equation}
This shows that the last terms of (\ref{eq:DtypeSP}) and (\ref{eq:SWreformulated}) indeed match upon making the identification $b = 4 \eta(\rho)^{18} \Delta^{(2)}$! The first parts also match as the different possible terms involve
\begin{equation}
	\theta_4^8, \theta_4^6 \theta_1^2, \theta_4^4 \theta_1^4, \theta_4^2 \theta_1^6, \theta_1^8,
\end{equation}
which can be written as products of elements of $H^0(L^4)_+$. This concludes our proof of the identification of the two curves (\ref{eq:DtypeSP}) and (\ref{eq:SWcurve}).

\section{Conclusions}

In this paper we proposed a novel construction of quantum states of M5 branes probing A-type and D-type singularities in terms of theta functions. We utilized powerful mathematical tools of theta functions to analyze different duality frames of M5 branes probing A-type singularities. In the D-type case an orbifold construction gives the quantum states of the corresponding LST. This formalism allows us then to elegantly derive Seiberg-Witten curves for torus compactifications of our LSTs which match with previously obtained results. One aspect we would like to emphasize here is that D-type quantum states as given for example in (\ref{eq:bigthetaX}) have a similar form as A-type states for the mass-less limit in a particular region in moduli space and thus there seem to be non-trivial relations between the two. Such relations have also recently been observed by \cite{Bastian:2018jlf} from a different point of view and it would be very interesting to explore this further.

From the point of view of geometric engineering the Little String Theories analyzed are obtained from F-theory compactifications on local elliptic Calabi-Yau three-folds. In this context the Seiberg-Witten curves we wirte down are mirror curves of the respective Calabi-Yau manifolds. Indeed, the curve equation given in (\ref{eq:LGcurve}) resembles very much mirror curves obtained from Landau-Ginzburg models, see for example \cite{Lerche:1989uy}, and from this point of view our $\mathcal{W}$ should have the interpretation of a superpotential (or be related to it). We are not dealing with typical Landau-Ginzburg models here, however, as our coordinates are theta functions and come with pre-defined relations between each other. Rather, one should think of our constructions in the context of the mathematical works \cite{Gross:2012du,Gross:2014fwa} where generalized notions of theta functions are introduced. It is in this generalized framework that we believe one can tackle the question of mirror curves for LSTs arising from M5 branes probing E-type singularities.  

Another aspect is that everything we have done in this paper is within a specific limit of the Omega background, namely we put our M5 branes on $T^2 \times_{\epsilon_1,\epsilon_2} \mathbb{R}^4$ and take the limit $\epsilon_1 = -\epsilon_2 \rightarrow 0$. Indeed this is the same limit taken in \cite{Haghighat:2017vch,Haghighat:2018dwe} when deriving the Seiberg-Witten curves using the formalism of the ``thermodynamic limit". It would be very interesting to repeat our analysis for the case of the full Omega background, i.e. without taking a specific limit. The corresponding quantum states will then be full BPS partition functions and should be related to the partition functions obtained in \cite{Haghighat:2013gba,Kim:2014dza,Haghighat:2014vxa,Gadde:2015tra,Kim:2017xan}. To perform such a computation, it might be important to switch to the framework of the blow-up equations for 6d SCFTs \cite{Gu:2018gmy} where it is reasonable to expect that different quantum states correspond to different fluxes of the $B$ field through the blow-up divisor of $\widehat{\mathbb{C}^2}$.

\acknowledgments
We would like to thank Michele Del Zotto, Guglielmo Lockhart, Hossein Movasati, Mauricio Romo, and Edward Witten for valuable discussions. The work of BH and RS was supported by the National Thousand-Young-Talents Program of China. BH would also like to thank the Simons Center for Geometry and Physics, the Institute for Advanced Study in Princeton and the Aspen Center for Physics for hospitality where part this work was carried out.

\end{document}